\documentclass[journal]{IEEEtran}

\usepackage{cite}
\usepackage{amsmath,amssymb,amsfonts,bm}
\usepackage{graphicx}
\usepackage{textcomp}
\usepackage{xcolor}
\usepackage{multicol}
\usepackage{booktabs}
\usepackage{multirow}
\usepackage{gensymb}
\usepackage{algorithm}

\usepackage{soul}
\usepackage{graphicx}
\usepackage{amsmath}
\usepackage{booktabs}
\usepackage{threeparttable}
\usepackage[hidelinks]{hyperref}

\begin{document}

\title{Secure and Lightweight Strong PUF Challenge Obfuscation with Keyed Non-linear FSR}

\author{\IEEEauthorblockN{Kleber Stangherlin, Zhuanhao Wu, Hiren Patel, Manoj Sachdev} \\
        \IEEEauthorblockA{ECE Department, University of Waterloo, Waterloo, ON N2L 3G1, Canada}\\
        \{khstangh, zhuanhao.wu, hiren.patel, msachdev\}@uwaterloo.ca
}

\markboth{Submitted to IEEE for possible publication. Copyright may be transferred. This version may no longer be accessible.}{}

\maketitle

\begin{abstract}
We propose a secure and lightweight key based challenge obfuscation for strong PUFs. Our architecture is designed to be resilient against learning attacks. Our obfuscation mechanism uses non-linear feedback shift registers (NLFSRs). Responses are directly provided to the user, without error correction or extra post-processing steps. We also discuss the cost of protecting our architecture against power analysis attacks with clock randomization, and Boolean masking. Security against learning attacks is assessed using avalanche criterion, and deep-neural network attacks. We designed a testchip in 65~nm CMOS. When compared to the baseline arbiter PUF implementation, the cost increase of our proposed architecture is 1.27x, and 2.2x when using clock randomization, and Boolean masking, respectively.
\end{abstract}


\section{Introduction}

Counterfeit integrated circuits (ICs) enter the supply chain from recycled, remarked, overproduced, cloned, or even out-of-spec, and defective parts~\cite{tutorialCounterfeit2014}. Physical unclonable functions (PUFs) offer a mechanism to help prevent counterfeiting~\cite{seminalArb2002}. Unlike traditional identification alternatives, two authentic ICs using PUFs have very low probability of producing the same output, regardless of their programmed memory content. The ideal PUF design is lightweight~\cite{seminalControlled2002}, secure against modeling attacks~\cite{seminalAttack2013}, and produces high entropy responses that are chip unique, and stable over various environmental conditions.

Previous works used composition of strong PUFs to obfuscate the internal challenge, but resilience to learning attacks is ultimately limited by the stability of responses~\cite{pufCpuf2014,pufPop2019,pufPopSi2022}. Error corrected strong PUFs require external helper data to cope with its large challenge space, but such alternative was shown to be insecure~\cite{pufEccVulnerable2014}. Recent works employ weak PUFs to generate an error corrected chip-unique secret, which is then used to obfuscate the external challenge~\cite{pufMPuf2018,pufSlate2019,pufGoingDeep2020}. The obfuscation algorithm must be lightweight, and secure against learning attacks. Moreover, manipulating external data (challenge) with sensitive information (secret key), requires counter-measures against power analysis attacks~\cite{seminalDpa1999}.

\begin{figure}[t]
    \centering
    \includegraphics[scale=1]{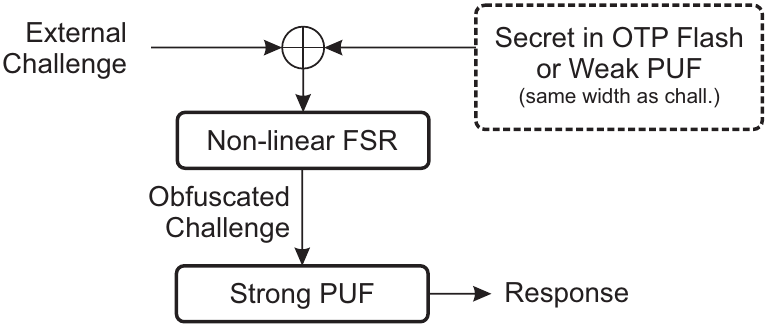}
    \caption{High-level view of our proposed key based challenge obfuscation architecture for strong PUFs.}
    \label{fig:arch}
\end{figure}

In this work we propose a secure and lightweight key based challenge obfuscation for strong PUFs. Our architecture is designed to be resistant against learning attacks, see Fig. \ref{fig:arch}. First, we XOR the external challenge with a secret key. The result is loaded into a non-linear feedback shift register (NLFSR), which is run for a number of cycles before the first evaluation (warm-up). The NLFSR state is then used as (obfuscated) challenge to evaluate the strong PUF. Responses are directly provided to the user---no error correction or post-processing is required. The secret key may be implemented with a one time programmable (OTP) flash, or a weak PUF.

Our key based challenge obfuscation architecture is designed to be resistant against learning attacks. We show that every bit of the obfuscated challenge meets the avalanche criterion~\cite{othersSac1985}. We also performed deep-neural network (DNN) experiments with an arbiter PUF (APUF) enhanced by our obfuscation technique. The DNN model failed to obtain generalized knowledge despite of the well known APUF vulnerabilities.

The mitigation of side-channel attacks is crucial for any key based challenge obfuscation. If the key is extracted, the strong PUF is exposed to attackers. Protecting hardware implementations against power analysis attacks is costly, but effective~\cite{mskFirst1999, scaRiscv2022}. We made careful design choices in our obfuscation architecture, such that side-channel mitigation techniques have minimum impact in cost. We also discuss the implementation of a clock randomization counter-measure against power analysis attacks.

We designed a 65~nm CMOS testchip with an APUF enhanced by our proposed challenge obfuscation architecture. Our design was submitted for fabrication, and its register transfer level (RTL) code is publicly available~\cite{resCodeKNL}. We provide detailed post-layout area results that explore the trade-off between side-channel attack resilience and cost.

\section{Challenge Obfuscation}

\subsection{Non-linear Feedback Shift Register (NLFSR)\label{sec:nlfsr}}

Non-linearity is a fundamental property for obfuscation algorithms. In the case of block ciphers, non-linear transformations are performed by substitution boxes (SBOXes). The implementation cost of SBOXes, however, is significant~\cite{pufIntel2020}. To achieve lightweight non-linear challenge transformation, we use NLFSRs. NLFSRs are deterministic digital circuits capable of non-linear state transitions. Unlike their linear counter-part, NLFSRs lack a solid mathematical representation. The sequence length is found using brute force methods, therefore, maximum length NLFSRs hardly exceed $2^{31}$~\cite{othersNlfsrPoly2014}.

\begin{figure}[t]
    \centering
    \includegraphics[scale=1]{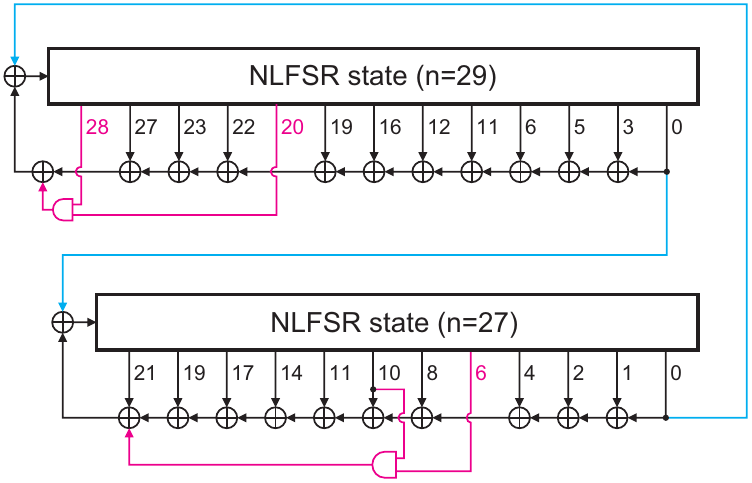}
    \caption{NLFSR arrangement used in challenge obfuscation architecture.}
    \label{fig:nlfsr}
\end{figure}

Our NLFSR design is shown in Fig. \ref{fig:nlfsr}. It uses a composition of two smaller NLFSRs, with 27 and 29~bits. Our feedback expressions have maximum length and were taken from~\cite{othersNlfsrPoly2014}. A 56~bit challenge is obtained by concatenating the state of both NLFSRs. To make both states dependent of one another, we XOR the lowest significant bit of each NLFSR with the next state logic of the other---a similar technique was used in the Trivium cipher~\cite{bookTrivium2008}. Registers are required to hold the challenge even if obfuscation is not used. Therefore, in addition to the control logic, the only overhead present in Fig. \ref{fig:nlfsr} are the logic gates that compute the next state.

Unless otherwise specified, we may refer to the 56~bit concatenated state simply as NLFSR state. Moreover, when clear from context, the term \textit{state} might be omitted.

\begin{algorithm}[t]
    \caption{Evaluate external challenge using an implementation without side-channel attack counter-measures.}
    \label{algo:eval}
    \vspace{0.1cm}\textbf{Assumptions:} secret key has been read from OTP flash, or weak PUF. Both secret key and external challenge are 56~bits.\vspace{0.1cm}
\begin{enumerate}
    \item XOR key with ext. challenge, and load result to NLFSR
    \item Run NLFSR for 112 cycles (for warm-up)
    \item Run NLFSR for 56 cycles (flush state)\label{algo:eval:flush}
    \item Evaluate strong PUF with NLFSR state as challenge
    \item Output the 1 bit response (no post-processing)
    \item If number of response bits is enough: \textit{return success}
    \item Otherwise: \textit{goto} step \ref{algo:eval:flush}
\end{enumerate}
\end{algorithm}

\subsection{Evaluation Algorithm (no counter-measures)}

The evaluation procedure for our proposed challenge obfuscation is listed in Algorithm \ref{algo:eval}. This algorithm is not yet protected against side-channel attacks. It assumes the secret key was read from OTP flash or weak PUF. Initially, the external challenge is XORed with secret key, and loaded in the NLFSR. The NLFSR is run for a total of 112 \textit{warm-up} cycles. Before each strong PUF evaluation, we \textit{flush} the previous NLFSR state, running it for 56 cycles.

More warm-up cycles enhances security, but increases latency. The number of warm-up cycles (112) was determined empirically to satisfy the avalanche criterion (see section~\ref{sec:aval}). Other NLFSR expression may require shorter, or longer warm-up periods to satisfy the avalanche criterion.

\begin{figure}[t]
    \centering
    \includegraphics[scale=1]{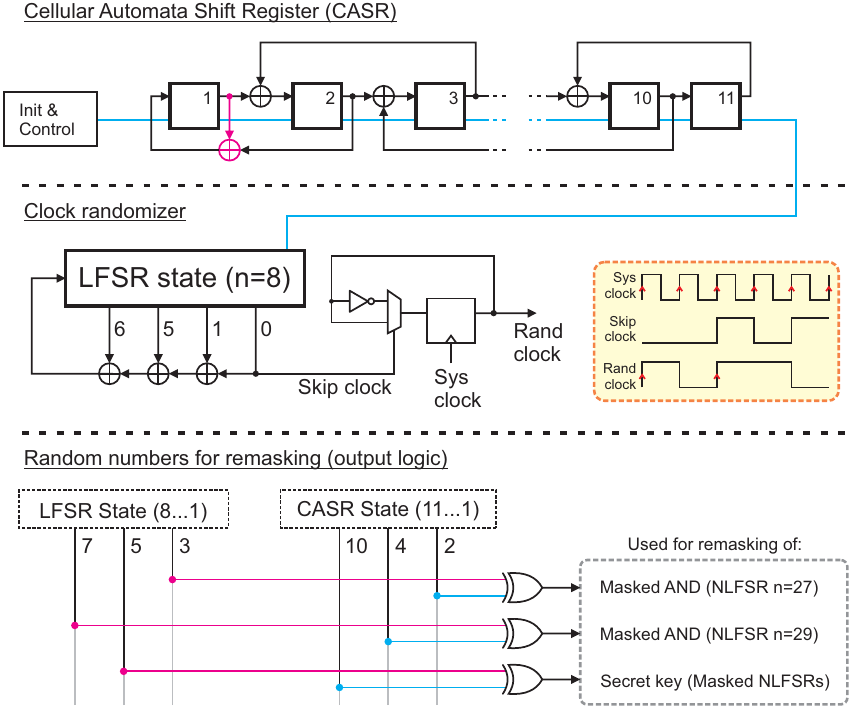}
    \caption{Clock randomization logic and pseudo random number generator (PRNG).}
    \label{fig:clkrnd}
\end{figure}

\subsection{Secret Key Storage}

The secret key storage may be implemented with one time programmable (OTP) flash, or weak PUF. Since uniqueness and unclonability properties are already provided by the strong PUF, using a weak PUF for secret key storage is possible, but adds extra complexity. One may argue that if a secret key is stored in flash, there is no need for a strong PUF. Such statement is inaccurate. For example, if the strong PUF is removed, over-produced chips (with blank OTP flash) can be programmed to behave alike any other device. That is not feasible when a strong PUFs is included in the design.

Other relevant considerations for both secret key storage options include protection against fault injection attacks, such as voltage glitches. Storing error correction data with the secret key, and performing multiple reads from memory for consistency checking was shown very effective in mitigating fault injections~\cite{faultBootloader2020}.

\section{Side-channel Attack Mitigation}

The power consumption can be used to extract secret information from unprotected devices. In particular, manipulating external data (challenge) with sensitive information (secret key) is vulnerable to power analysis attacks~\cite{seminalDpa1999}. This section discusses counter-measures to mitigate such risks.

\subsection{Random Number Generator (RNG)}

\begin{figure*}[t]
    \centering
    \includegraphics[scale=1]{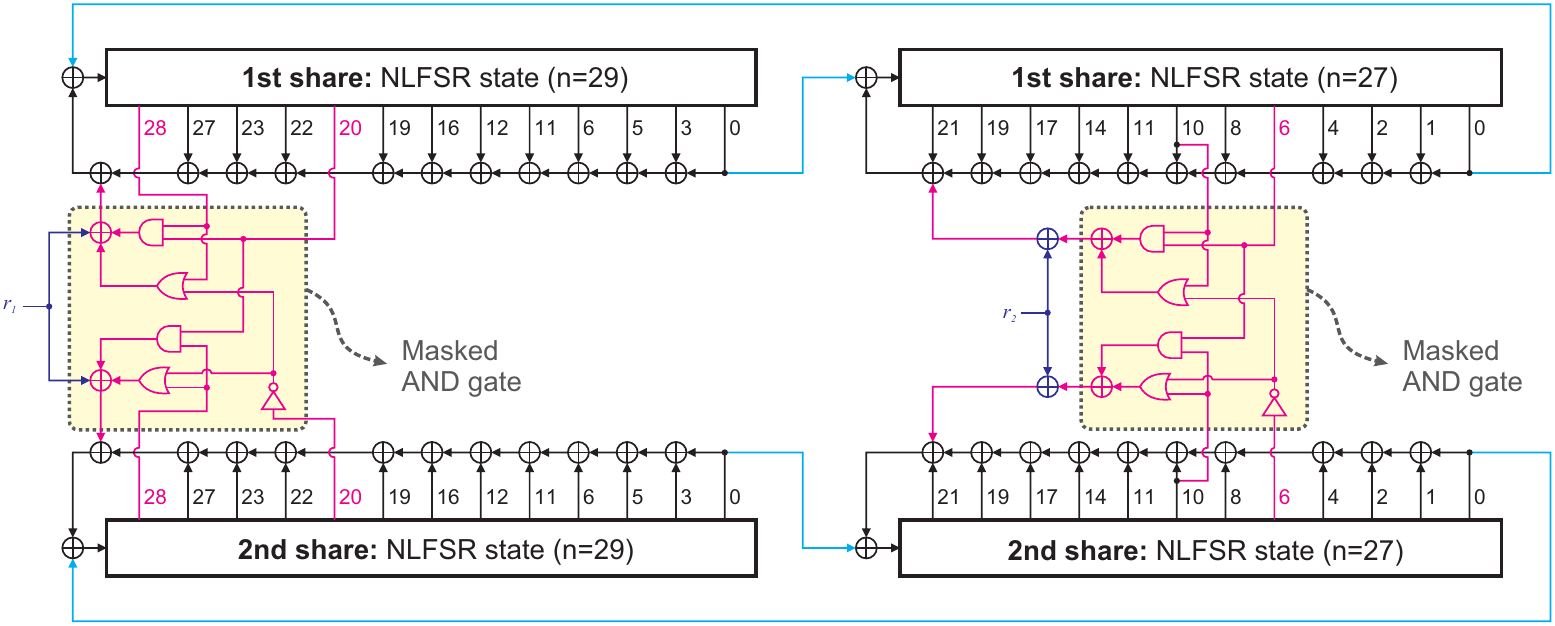}
    \caption{Boolean masked implementation of the NLFSR used for challenge obfuscation. Random numbers $r_1$ and $r_2$ are outputs from the PRNG.}
    \label{fig:masked}
\end{figure*}

Random numbers are necessary to implement the counter-measures described in this section. Our pseudo random number generator (PRNG) is based on \cite{scaRiscv2022}, and is shown in Fig. \ref{fig:clkrnd}. It uses an LFSR and a cellular automata shift register (CASR), with outputs derived from their combined (XORed) state. The PRNG has three outputs, each of them generates a new random number every clock cycle. LFSR and CASR have maximum sequence length, with expressions taken from \cite{othersLfsrPoly1973}, and \cite{rngHyridCa1995}. Because their cycle length is relatively prime, the total cycle length of the output is close to $2^{19}$. The seeding of LFSR/CASR states is performed using the available strong PUF. For that, we identify a challenge with unstable responses during enrollment, and store it in non-volatile memory. This challenge is repeatedly evaluated to generate random bits that initialize the LFSR/CASR states.

\subsection{Clock Randomization}

Power analysis attacks extract the key over a large number of recorded power traces. In each trace, the device manipulates distinct external data using the same key. The effectiveness of power attacks is higher when traces are aligned in time. The clock randomizer produces an irregular clock waveform, which will randomly skip clock edges. Our obfuscation architecture uses the randomized clock output. The circuit is shown in Fig. \ref{fig:clkrnd}, and it may be seen as a conditional clock divider. The \textit{skip clock} signal is derived from the 8~bit LFSR, therefore, the randomized clock waveform pattern repeats every 255 cycles, allowing authentication with predictable performance.

\subsection{Boolean Masking of the NLFSRs}

Boolean masking uses random numbers to split each value into two shares that are (ideally) uncorrelated to the original value. For example, the shared representation of $x$ is $(x_1, x_2)$, which are computed as $x_1 = x \oplus r$, and $x_2 = r$, where $r$ is a random number. The original value is recovered by XORing the shares, therefore, $x = x_1 \oplus x_2$. Physical probing of both shares is necessary to inspect the original data. When operations are performed with shared data, the power consumption is, in theory, uncorrelated to the data being processed. Moreover, if the secret key is stored in two shares, its plain-text value is never exposed.

Converting conventional single share circuits to operate with multiple shares of data differs for linear and non-linear operations. Linear operations, like XOR, shifts, and permutations, are simply applied to both shares without any changes. Non-linear operations, however, require a replacement circuit that will compute the shared outputs without disclosing the original value. We redesigned the NLFSRs described in section \ref{sec:nlfsr} to run using multiple shares. The resulting circuit is shown in Fig. \ref{fig:masked}. The number of state registers doubles to accommodate both shares of all values. Since XORs and shifts are linear operations, the NLFSR design remains mostly unchanged, except for the AND gates, which are replaced by their masked counterpart. The expression for the masked AND was taken from \cite{mskBiru2017}. In fact, during the design of our challenge obfuscation architecture, we intentionally selected NLFSR expressions with a small number of non-linear gates to reduce the cost and complexity of masked implementations. For example, glitches are a well known source of information leakage in masked non-linear logic~\cite{mskGlitches2005}, but our original design has both AND inputs driven by registers---the best practice was already implemented.

\begin{algorithm}[t]
    \caption{Evaluate  external challenge using an implementation with clock randomization, and Boolean masking.}
    \label{algo:seceval}
    \vspace{0.1cm}\textbf{Assumptions:} secret key has been read from OTP flash, or weak PUF in two shares of 56~bits each. External challenge is 56~bits, with second share being all zeros.\vspace{0.1cm}
\begin{enumerate}
    \item Using the challenge with unstable responses, (serially) seed the PRNG with random numbers (8~+~11~=~19~bits)
    \item Enable the clock randomizer
    \item Using the PRNG output, (serially) seed the NLFSR with the same 56~bit random number in both shares
    \item XOR the NLFSR content with the key, and load result back to NLFSR (for key remasking) 
    \item Run the NLFSR for 128 cycles (for time misalignment)
    \item XOR ext. challenge with NLFSR, and load result back to NLFSR
    \item Run the NLFSR for 112 cycles (for warm-up)
    \item Run NLFSR for 56 cycles (flush state)\label{algo:seceval:flush}
    \item XOR the NLFSR shares and evaluate strong PUF with the unmasked state (obfuscated challenge)
    \item Output the 1 bit response (no post-processing needed)
    \item If number of response bits is enough: \textit{return success}
    \item Otherwise: \textit{goto} step \ref{algo:seceval:flush}
\end{enumerate}
\end{algorithm}

Other design details include remasking the AND gate at every cycle, which is done by XORing fresh random numbers, $r_1$ and $r_2$, with both shares of the masked AND output. Remasking helps remove possible correlations between the original data and the shared values.

When the NLFSR warm-up cycles are completed, their shared state is XORed to obtain the unmasked, obfuscated challenge, necessary for evaluation. It is very important that this XOR operation is only performed after completion of all warm-up cycles. In other words, the XOR inputs must be gated during warm-up to avoid information leakage from the toggling XOR outputs.

\subsection{Evaluation Algorithm (with counter-measures)}

The evaluation procedure for the masked implementation is listed in Algorithm \ref{algo:seceval}. Similarly to Algorithm \ref{algo:eval}, it assumes that the secret key was read from OTP flash, or weak PUF. However, the key is expected in two shares of 56~bits each. First, the PRNG is seeded and the clock randomizer is enabled. The NLFSR is then seeded with random numbers from the PRNG. Both shares must be loaded with the same random number. The secret key is then XORed with the NLFSR state, which essentially performs a remasking operation of the key shares. Next, the NLFSR is run for 128 cycles to allow random delay insertion by the clock randomizer. The external challenge is then XORed with current NLFSR state, where the second share is considered all zeros. This will not leak information since the NLFSR content is already masked with fresh random numbers. The remaining steps are analogous to Algorithm \ref{algo:eval}, with the extra requirement of XORing the NLFSR shares to unmask the obfuscated challenge before each strong PUF evaluation.

\section{Testchip Implementation\label{sec:testchip}}

\begin{figure}[t]
    \centering
    \includegraphics[scale=0.75]{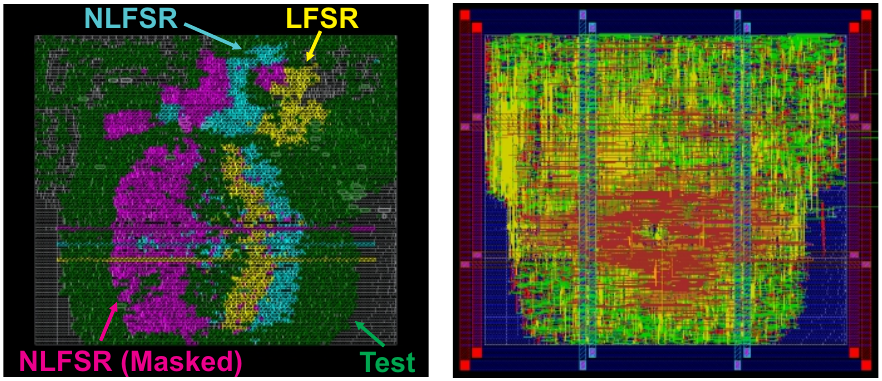}
    \caption{Testchip placement of standard-cells and layout in 65~nm CMOS.}
    \label{fig:layout}
\end{figure}

To accurately assess the effectiveness of our counter-measures against power analysis attacks, we designed a 65~nm CMOS testchip. Our design was submitted for fabrication, and the RTL code is publicly available~\cite{resCodeKNL}. Table \ref{tab:area} shows the post-layout area results for the three implementations included in the testchip. The number of instances, and area are listed for each implementation, and its components. Area for test logic is not reported. All implementations use 7 repeated evaluations for enhanced response stability~\cite{pufPopSi2022}. The 56-APUF is a custom designed arbiter PUF with 56 delay stages, see~\cite{pufPopSi2022} for APUF circuit details. The clock randomizer and CASR were split in two different design blocks. The placement of standard-cells and layout is shown in Fig. \ref{fig:layout}.

The 56-LFSR-APUF uses an APUF with challenge stored/unrolled by an LFSR. It represents the smallest viable authentication system that can be built using an arbiter PUF. It uses an area of 1583 ND2, which we define as our comparison baseline. The 56-NLFSR-APUF implements our challenge obfuscation architecture, without Boolean masking. Nevertheless, this implementation is not completely unprotected against side-channel attacks. It includes a clock randomizer instance, which represents 5\% of the used area. The overall area of the 56-NLFSR-APUF implementation is 27\% (1.27x) larger than the baseline. Finally, the 56-NLFSR-APUF (Masked) denotes the fully masked implementation, with clock randomization. The area for this implementation is 120\% (2.2x) larger than the baseline. Results reported in~\cite{scaRiscv2022} suggest that clock randomization alone delivers 2260x increase in resilience against power analysis attacks. An even greater increase, of 17330x, is achieved when clock randomization is combined with Boolean masking.

\begin{table}[t]
    \centering
    \caption{Testchip area results in 65~nm CMOS (post-layout).}
    \label{tab:area}
    \begin{threeparttable}
        \begin{tabular}{@{}lrrr@{}}
\toprule
\textbf{}   & \multicolumn{1}{l}{\textbf{\# Inst.}} & \multicolumn{1}{l}{\textbf{Area ($\mu m^2$)}} & \multicolumn{1}{l}{\textbf{Area (ND2)}} \\ \midrule
\textbf{56-LFSR-APUF}                       & \textbf{490}      & \textbf{2279}   & \textbf{1583}     \\
\hspace{0.5cm}LFSR                          & 152               & 754             & 524               \\
\hspace{0.5cm}56-APUF                       & 1                 & 332             & 231               \\
\hspace{0.5cm}Control logic \& buffers      & 337               & 1193            & 829              \\ \midrule
\textbf{56-NLFSR-APUF}                      & \textbf{717}      & \textbf{2904}   & \textbf{2017}     \\
\hspace{0.5cm}NLFSR                         & 186               & 854             & 593               \\
\hspace{0.5cm}Clock randomizer              & 24                & 148             & 103               \\
\hspace{0.5cm}56-APUF                       & 1                 & 332             & 231               \\
\hspace{0.5cm}Control logic \& buffers      & 506               & 1570            & 1090              \\ \midrule
\textbf{56-NLFSR-APUF (Masked)}             & \textbf{1448}     & \textbf{5016}   & \textbf{3483}     \\
\hspace{0.5cm}Masked NLFSR                  & 494               & 2057            & 1428              \\
\hspace{0.5cm}Clock randomizer              & 25                & 148             & 103               \\
\hspace{0.5cm}CASR                          & 65                & 237             & 165               \\
\hspace{0.5cm}56-APUF                       & 1                 & 332             & 231               \\
\hspace{0.5cm}Control logic \& buffers      & 863               & 2243            & 1557              \\ \bottomrule
\end{tabular}

        \begin{tablenotes}[para,flushleft]
            Notes: area associated to the secret key storage/read is not included. Area is reported in $\mu m^2$ and normalized by NAND2.
        \end{tablenotes}
    \end{threeparttable}
\end{table}

\section{Security Assessment}

\subsection{Avalanche Criterion\label{sec:aval}}

As defined in~\cite{othersSac1985}, if a cryptographic function is to satisfy the strict avalanche criterion (SAC), then, each output bit should change with a probability of one half, whenever a single input bit is complemented. We assess the avalanche criterion from the perspective of our obfuscation algorithm---input is the external challenge, and output is the NLFSR state after (112 + 56) cycles. Our experiment used 10000 unique external challenges, each of them was run twice, with a single toggled bit between runs. The secret key is randomized at the beginning and remains unchanged. 

\begin{figure}[t]
    \centering
    \includegraphics[scale=1]{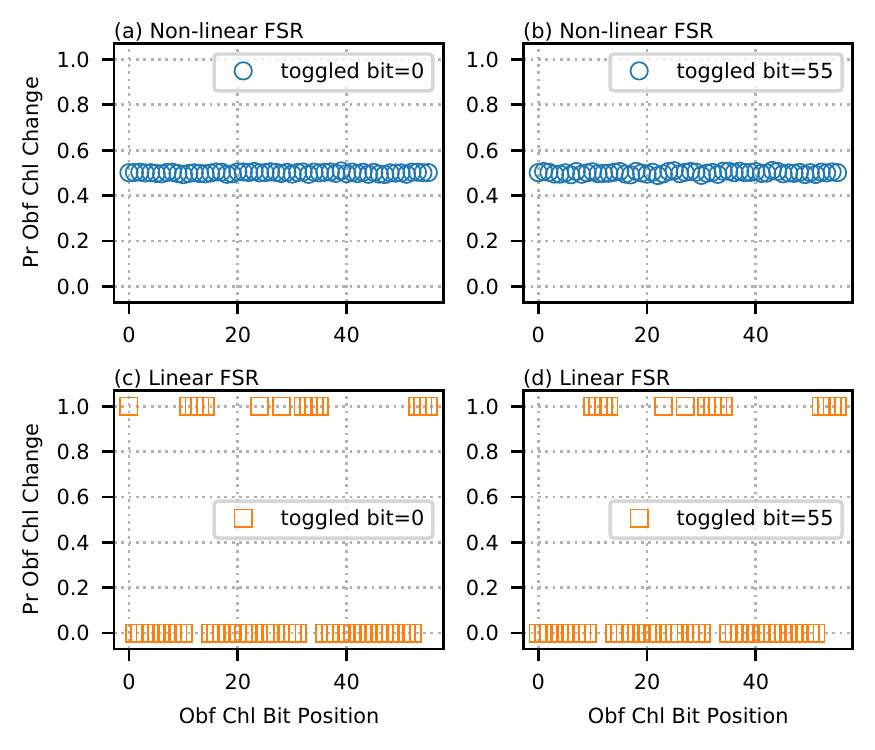}
    \caption{Probability of change in NLFSR/LFSR state when a single bit of the external challenge changes.}
    \label{fig:prout}
\end{figure}

Fig. \ref{fig:prout} (a) and (b) show that toggling a single input bit causes a widespread (avalanche) effect in the NLFSR state, where each obfuscated challenge bit has an estimated 50\% probability of changing. Same behavior was observed when other bit positions are toggled. For comparison, we replaced the NLFSR by a 56~bit linear feedback shift register (LFSR) of maximum sequence length (same as our baseline implementation discussed in section \ref{sec:testchip}). The LFSR experiment also evaluates using a secret key, with output taken after (112 + 56) cycles. Results for the LFSR are reported in Fig. \ref{fig:prout} (c) and (d), showing that the effects of a single toggled input bit on the final LFSR state are deterministic, that is, either 0\%, or 100\%. Therefore, LFSR based challenge obfuscation fails to meet the avalanche criterion.

\subsection{DNN Attacks}

Deep-neural networks (DNNs) are capable of learning complex PUF structures, without a mathematical model of the PUF being modelled. We performed experiments using a challenge obfuscated arbiter PUF, and a 12-layer DNN architecture similar to~\cite{attackDNN2019}. The input and output layers have 56, and 2 units, with 2000 units in hidden layers. The DNN was trained for 72 hours using 10 million CRPs, with a resulting accuracy equivalent to the PUF uniformity bias, which was 56\% in our experiment. Therefore, the DNN model failed to obtain generalized learning on the challenge obfuscated arbiter PUF.

\section{Conclusion}

We demonstrated the design of a lightweight key based challenge obfuscation for strong PUFs. We addressed security for both learning, and power analysis attacks. Future work shall use the fabricated testchip to assess the effectiveness of our side-channel attack counter-measures.

\newpage

\bibliographystyle{plain}
\bibliography{knl}

\begin{thebibliography}{10}

\bibitem{mskBiru2017}
A~Biryukov, D~Dinu, YL~Corre, and A~Udovenko.
\newblock Optimal first-order boolean masking for embedded {IoT} devices.
\newblock In {\em Intl. Conf. on Smart Card Research and Advanced
  Applications}, pages 22--41. Springer, 2017.

\bibitem{rngHyridCa1995}
K~Cattell and S~Zhang.
\newblock Minimal cost one-dimensional linear hybrid cellular automata of
  degree through 500.
\newblock {\em JET}, 6(2):255--258, 1995.

\bibitem{othersNlfsrPoly2014}
P~Dabrowski, G~Labuzek, T~Rachwalik, and J~Szmidt.
\newblock Searching for nonlinear feedback shift registers with parallel
  computing.
\newblock {\em Information Processing Letters}, 114(5):268--272, 2014.

\bibitem{pufEccVulnerable2014}
J~Delvaux and I~Verbauwhede.
\newblock Key-recovery attacks on various {RO PUF} constructions via helper
  data manipulation.
\newblock In {\em DATE}, pages 1--6. IEEE, 2014.

\bibitem{seminalControlled2002}
Blaise Gassend, Dwaine Clarke, Marten Van~Dijk, and Srinivas Devadas.
\newblock Controlled physical random functions.
\newblock In {\em Computer Security Applications Conference}, pages 149--160.
  IEEE, 2002.

\bibitem{seminalArb2002}
Blaise Gassend, Dwaine Clarke, Marten Van~Dijk, and Srinivas Devadas.
\newblock Silicon physical random functions.
\newblock In {\em Conference on Computer and Communications Security}, pages
  148--160. ACM, 2002.

\bibitem{mskFirst1999}
L~Goubin and J~Patarin.
\newblock {DES} and differential power analysis the ``duplication" method.
\newblock In {\em CHES}, pages 158--172. Springer, 1999.

\bibitem{tutorialCounterfeit2014}
U~Guin, K~Huang, D~DiMase, JM~Carulli, M~Tehranipoor, and Y~Makris.
\newblock Counterfeit integrated circuits: A rising threat in the global
  semiconductor supply chain.
\newblock {\em Proceedings of the IEEE}, 102(8):1207--1228, 2014.

\bibitem{pufGoingDeep2020}
M~Khalafalla, MA~Elmohr, and C~Gebotys.
\newblock Going deep: Using deep learning techniques with simplified
  mathematical models against {XOR} {BR} and {TBR} {PUFs} (attacks and
  countermeasures).
\newblock In {\em HOST}, pages 80--90. IEEE, 2020.

\bibitem{attackDNN2019}
M~Khalafalla and C~Gebotys.
\newblock {PUFs} deep attacks: Enhanced modeling attacks using deep learning
  techniques to break the security of double arbiter {PUFs}.
\newblock In {\em DATE}, pages 204--209, 2019.

\bibitem{seminalDpa1999}
P~Kocher, J~Jaffe, and B~Jun.
\newblock Differential power analysis.
\newblock In {\em Annual Intl. Cryptology Conf.}, pages 388--397. Springer,
  1999.

\bibitem{pufMPuf2018}
Q~Ma, C~Gu, N~Hanley, C~Wang, W~Liu, and M~O'Neill.
\newblock A machine learning attack resistant multi-{PUF} design on {FPGA}.
\newblock In {\em ASP-DAC}, pages 97--104. IEEE, 2018.

\bibitem{mskGlitches2005}
S~Mangard, T~Popp, and BM~Gammel.
\newblock Side-channel leakage of masked {CMOS} gates.
\newblock In {\em Cryptographers Track at the RSA Conf.}, pages 351--365.
  Springer, 2005.

\bibitem{bookTrivium2008}
M~Robshaw and O~Billet.
\newblock {\em New stream cipher designs: the eSTREAM finalists}, volume 4986.
\newblock Springer, 2008.

\bibitem{seminalAttack2013}
Ulrich R{\"u}hrmair, Jan S{\"o}lter, Frank Sehnke, Xiaolin Xu, Ahmed Mahmoud,
  Vera Stoyanova, Gideon Dror, J{\"u}rgen Schmidhuber, Wayne Burleson, and
  Srinivas Devadas.
\newblock {PUF} modeling attacks on simulated and silicon data.
\newblock {\em Transactions on Information Forensics and Security},
  8(11):1876--1891, 2013.

\bibitem{pufCpuf2014}
DP~Sahoo, S~Saha, D~Mukhopadhyay, RS~Chakraborty, and H~Kapoor.
\newblock Composite {PUF}: A new design paradigm for physically unclonable
  functions on {FPGA}.
\newblock In {\em HOST}, pages 50--55. IEEE, 2014.

\bibitem{othersLfsrPoly1973}
W~Stahnke.
\newblock Primitive binary polynomials.
\newblock {\em Mathematics of computation}, 27(124):977--980, 1973.

\bibitem{scaRiscv2022}
K~Stangherlin and M~Sachdev.
\newblock Design and implementation of a secure risc-v microprocessor.
\newblock {\em arXiv preprint arXiv:2205.05095}, 2022.

\bibitem{resCodeKNL}
K~Stangherlin, Z~Wu, H~Patel, and M~Sachdev.
\newblock Secure and lightweight challenge obfuscation with keyed non-linear
  feedback shift register ({RTL} code).
\newblock \url{https://github.com/cdrlabs-waterloo/2022-07-22}.

\bibitem{pufPopSi2022}
K~Stangherlin, Z~Wu, H~Patel, and M~Sachdev.
\newblock Design exploration and security assessment of puf-on-puf
  implementations.
\newblock {\em arXiv preprint arXiv:2206.11840}, 2022.

\bibitem{pufIntel2020}
V~Suresh, R~Kumar, M~Anders, H~Kaul, V~De, and S~Mathew.
\newblock A 0.26\% {BER}, 10 28 challenge-response machine-learning resistant
  strong-{PUF} in 14nm {CMOS} featuring stability-aware adversarial challenge
  selection.
\newblock In {\em VLSI Symp.}, pages 1--2. IEEE, 2020.

\bibitem{faultBootloader2020}
J~Van~den Herrewegen, D~Oswald, FD~Garcia, and Q~Temeiza.
\newblock Fill your boots: Enhanced embedded bootloader exploits via fault
  injection and binary analysis.
\newblock {\em TCHES}, pages 56--81, 2021.

\bibitem{pufSlate2019}
W-C Wang, Y~Yona, Y~Wu, SN~Diggavi, and P~Gupta.
\newblock Slate: a secure lightweight entity authentication hardware primitive.
\newblock {\em TIFS}, 15:276--285, 2019.

\bibitem{othersSac1985}
AF~Webster and Stafford~E Tavares.
\newblock On the design of {S}-boxes.
\newblock In {\em Conf. on the theory and application of cryptographic
  techniques}, pages 523--534. Springer, 1985.

\bibitem{pufPop2019}
Z~Wu, H~Patel, M~Sachdev, and MV~Tripunitara.
\newblock Strengthening {PUFs} using composition.
\newblock In {\em ICCAD}, pages 1--8. IEEE, 2019.

\end{thebibliography}

\end{document}